\documentclass[secnumarabic, twocolumn, graphics,floatfix, superscriptaddress,tightenlines, aps, prb]{revtex4-2}
\usepackage[dvips]{graphicx}
\usepackage{amssymb,amsmath}
\usepackage{dcolumn}
\usepackage{bm}
\usepackage{color}
\usepackage[colorlinks=true, allcolors=blue]{hyperref}
\usepackage{multirow}
\usepackage{tabularx}

\newcommand{\cMK}[1]{\textcolor{blue}{{\bf MK:} #1}}
\newcommand{\cD}[1]{\textcolor{red}{{\bf Dima:} #1}}

\begin{document}

\title{Hyperfine interaction of electrons and holes with nuclei probed by optical orientation in MAPbI$_3$ perovskite crystals}

\author{Mladen Kotur}
\affiliation{Experimentelle Physik 2, Technische Universit\"{a}t Dortmund, 44227 Dortmund, Germany}
\author{Nataliia E. Kopteva}
\affiliation{Experimentelle Physik 2, Technische Universit\"{a}t Dortmund, 44227 Dortmund, Germany}
\author{Dmitri R. Yakovlev}
\affiliation{Experimentelle Physik 2, Technische Universit\"{a}t Dortmund, 44227 Dortmund, Germany}
\author{Bekir~Turedi}
\affiliation{Laboratory of Inorganic Chemistry, Department of Chemistry and Applied Biosciences,  ETH Z\"{u}rich, CH-8093 Z\"{u}rich, Switzerland}
\affiliation{Laboratory for Thin Films and Photovoltaics, Empa-Swiss Federal Laboratories for Materials Science and Technology, CH-8600 D\"{u}bendorf, Switzerland}
\author{Maksym~V.~Kovalenko}
\affiliation{Laboratory of Inorganic Chemistry, Department of Chemistry and Applied Biosciences,  ETH Z\"{u}rich, CH-8093 Z\"{u}rich, Switzerland}
\affiliation{Laboratory for Thin Films and Photovoltaics, Empa-Swiss Federal Laboratories for Materials Science and Technology, CH-8600 D\"{u}bendorf, Switzerland}
\author{Manfred Bayer}
\affiliation{Experimentelle Physik 2, Technische Universit\"{a}t Dortmund, 44227 Dortmund, Germany}
\affiliation{Research Center FEMS, Technische Universit\"at Dortmund, 44227 Dortmund, Germany}

\date{\today}

\begin{abstract}

Optical orientation of electron and hole spins by circularly polarized light is investigated for MAPbI$_3$ single crystals. The Hanle and polarization recovery effects measured in transverse and longitudinal magnetic fields, respectively, evidence the hyperfine interaction with nuclear spins as the main factor determining the spin dynamics of charge carriers at cryogenic temperatures. The parameters of the nuclear spin fluctuations within the carrier localization volume are evaluated. Dynamic polarization of the nuclear spins is demonstrated by the Overhauser field reaching $5$~mT for acting on the electrons and $-30$~mT for acting on the holes.

\end{abstract}

 \maketitle

\section{Introduction}
\label{sec:intro}
Lead halide perovskite semiconductors have become a focus of current research due to their technological success and promising applications in photovoltaics and optoelectronics~\cite{Vardeny2022_book,Vinattieri2021_book}. Their band structure, which considerably differs from conventional III-V and II-VI semiconductors, makes perovskite semiconductors interesting as testbed for basic research in spin physics and for applications exploiting spin-dependent phenomena in spintronics and quantum information technologies.

Optical spin orientation of excitons and charge carriers is a key phenomenon in the spin physics of semiconductors~\cite{meier1984optical,dyakonov2017}, which is used as a tool for several magneto-optical techniques. The classical approach uses circularly-polarized photoluminescence for the detection of the spin polarization of excitons or charge carriers photogenerated by circularly-polarized laser light. Both continuous-wave and pulsed excitation can be used, in the latter case time-resolved detection can be implemented. Using external magnetic fields enriches the potential of the technique. The optical orientation technique provides detailed information on exciton and carrier spin dynamics, as well as on their hyperfine interaction with the nuclear spin system~\cite{OOChapter5,SPSDyakonov}. It can also be used to dynamically polarize the nuclear spin system via its interaction with optically oriented carriers, as well as to read the state and dynamics of the nuclear polarization. A broad spectrum of spin phenomena has been disclosed and addressed by the  optical orientation  in conventional semiconductors.

Recently, the optical orientation of excitons and charge carriers has been studied in lead halide perovskite single crystals (CsPbBr$_3$, FAPbBr$_3$, MAPbI$_3$, and  FA$_{0.9}$Cs$_{0.1}$PbI$_{2.8}$Br$_{0.2}$) \cite{kopteva2023giant,kudlacik2024optical,Kopteva_2025OOX,kopteva2025optical}. A very high degree of optical orientation has been found, reaching 85\% for excitons and 60\% for carriers. The degree is robust with respect to detuning the excitation energy by more than 300 meV from the exciton resonance, which confirms the spatial inversion symmetry in perovskite semiconductors of different crystal symmetry (cubic, tetragonal, and orthorhombic) so that the Dyakonov-Perel spin relaxation mechanism is suppressed. 

The strong hyperfine interaction of the carrier spins with the nuclear spin system was first observed in perovskite crystals using time-resolved Kerr rotation, namely through dynamic nuclear polarization (DNP)~\cite{belykh2019coherent,kirstein2022lead,kirstein2022mapi}. Nuclear spin fluctuations play an important role in the spin dynamics of localized electrons and holes at cryogenic temperatures, as we showed for FA$_{0.9}$Cs$_{0.1}$PbI$_{2.8}$Br$_{0.2}$ crystals using the Hanle and polarization recovery effects~\cite{kudlacik2024optical,Kotur2026_FAPI}. The hyperfine interaction in lead halide perovskites was theoretically considered in Ref.~\cite{Kotur2026_FAPI}, concluding that for holes it is dominated by the $^{207}$Pb isotopes with spin $I=1/2$ and for electrons by the $^{207}$Pb isotopes and by the $^{127}$I isotopes with spin $I=5/2$.

In this paper, we use the optical orientation technique with continuous-wave excitation to study the effects of the hyperfine interaction between the nuclear spin system and localized electrons and holes in MAPbI$_3$ single crystals.

\section{Samples and experimental setup}
\label{sec:experiment}

We investigate a single crystal of methylammonium lead iodide (MAPbI$_3$) with a thickness of 30~$\mu$m (sample M2-7), prepared using the inverse temperature crystallization technique ~\citep{nazarenko2017single, chen2019single, alsalloum2020low}. At ambient temperature, the MAPbI$_3$ single crystal exhibits a tetragonal structure, with the [001] axis oriented perpendicular to the surface. When cooled to cryogenic temperatures, the crystal undergoes a phase transition at about $160^{\circ}$C to an orthorhombic structure.

The sample is placed in the variable temperature insert of a helium bath cryostat, where the temperature can be varied from 1.6~K up to 15~K. For temperatures above 4.2~K, the sample is cooled by a flow of helium gas, whereas at lower temperatures it is in direct contact with liquid helium. To measure photoluminescence (PL), the sample is excited using a continuous-wave Ti:Sapphire laser with photon energy of $E_{\rm exc} = 1.722$~eV, set above the MAPbI$_3$ band gap of $E_g = 1.652$~eV~\cite{kopteva2025optical}. Note that the exciton binding energy in MAPbI$_3$ is 16~meV~\cite{galkowski2016determination}.
The propagation direction of the laser beam given by the wave vector $\mathbf{k}$ is oriented along the [001] crystallographic axis, $\mathbf{k}\parallel[001]$. The laser is circularly polarized, with the helicity alternating between $\sigma^+$ and $\sigma^-$ at 400~kHz by an electro-optical modulator (EOM) driven with a pulse generator providing the trigger signal required to achieve the modulation frequency. Alternatively, constant $\sigma^+$  polarization of excitation is employed by replacing the EOM with a quarter-wave ($\lambda/4$) plate. 

The photoluminescence is collected in reflection geometry and analyzed with a 0.5-m spectrometer coupled to an avalanche photodiode. A two-channel gated photon counter, synchronized with the EOM, enables simultaneous measurement of the $I^{++}$ and $I^{-+}$ components of the $\sigma^+$-polarized photoluminescence under modulated $\sigma^+/\sigma^-$ excitation. A polarization analyzer in the detection path, composed of a $\lambda/4$ wave plate and a Glan–Thompson prism, enables distinction of the circularly polarized components of the emitted light. The degree of optical orientation is defined as
\begin{equation}
P_{\rm oo}=\frac{I^{++}-I^{-+}}{I^{++}+I^{-+}}.
\label{eq:optical_orientation_degree_modulated}
\end{equation}
Alternatively, under constant $\sigma^+$ excitation with a $\lambda/4$ plate in the excitation path, the degree of optical orientation is given by
\begin{equation}
P_{\rm oo}=\frac{I^{++}-I^{+-}}{I^{++}+I^{+-}}.
\label{eq:optical_orientation_degree_constant}
\end{equation}
Here, $I^{++}$ and $I^{+-}$ denote the photoluminescence intensities in $\sigma^+$ and $\sigma^-$ polarization, respectively. An external electromagnet, which can be rotated, provides a magnetic field $\textbf{B}$ with a strength of up to 125~mT. The field is applied in different geometries with respect to the $k$-vector: Faraday geometry ($\textbf{B}_{\rm F} \parallel \textbf{k}$) and Voigt geometry ($\textbf{B}_{\rm V} \perp \textbf{k}$).

\begin{figure}[hbt]
\center{\includegraphics{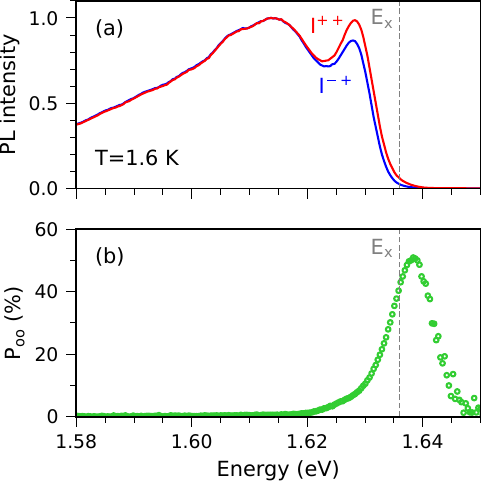}}
\caption{(a) Polarized photoluminescence spectra measured on a MAPbI$_3$ crystal at $T = 1.6$~K. The laser helicity is modulated between $\sigma^+$ and $\sigma^-$ at 400~kHz, while the detection polarization is fixed at $\sigma^+$. Laser photon energy $E_{\rm exc}=1.722$~eV with power $P=18$~W/cm$^2$. The dashed gray line indicates the free exciton energy, $E_{\rm X}=1.636$~eV, measured in reflectivity, see Ref.~\citep{kopteva2025optical}. (b) Spectral dependence of the optical orientation degree, evaluated according to Eq.~\eqref{eq:optical_orientation_degree_modulated} from the spectra shown in panel (a).}
\label{fig:spectrum}
\end{figure}

The photoluminescence (PL) spectrum of the MAPbI$_3$ single crystal, measured for  modulated $\sigma^+/\sigma^-$ excitation, is shown in Fig.~\ref{fig:spectrum}(a). It has a maximum at 1.628~eV that originates from the recombination of spatially separated localized carriers (electrons and holes)~\citep{wright2017band, dequilettes2019charge, kirstein2022mapi}. Other mechanisms, such as polaron formation~\citep{dequilettes2019charge}, carrier trapping and detrapping~\citep{chirvony2018trap}, and extended carrier diffusion~\citep{bercegol2018slow} may also contribute. The maximum of optical orientation of about 50\% is observed at 1.638~eV, see Fig.~\ref{fig:spectrum}(b), close to the exciton resonance at $E_{\rm X}=1.636$~eV~\citep{kopteva2025optical}. The spin signal is composed of contributions from electrons and holes as well as excitons, resulting from the overlap of the recombination of localized carriers with the exciton emission at  specific energies~\citep{kudlacik2024optical}.

\begin{figure*}[hbt]
\center{\includegraphics{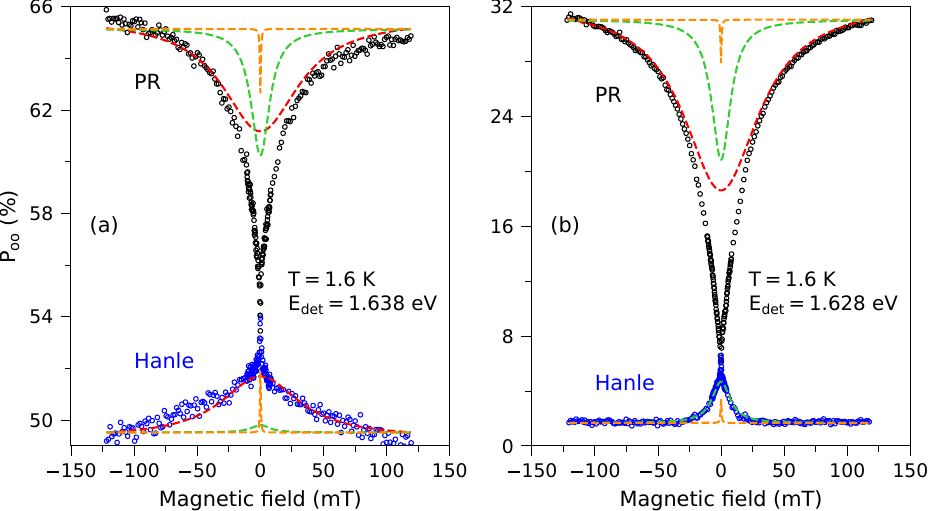}}
\caption{Hanle and polarization recovery curves measured in Voigt or Faraday
geometry at $T=1.6$~K for (a) $E_{\rm det}=1.638$~eV and (b) 1.628~eV photon detection energy. The laser light helicity is modulated at 400~kHz, with $E_{\rm exc} = 1.722$~eV and $P = 18$~W/cm$^2$. The dashed lines show the individual contributions of strongly localized holes (red), electrons (green), and weakly localized holes (orange). Their parameters are given in Table~\ref{tab:comparison_spectral}.}  
\label{fig:Hanle_PRC_Spectral}
\end{figure*}

\begin{figure*}[hbt]
\center{\includegraphics{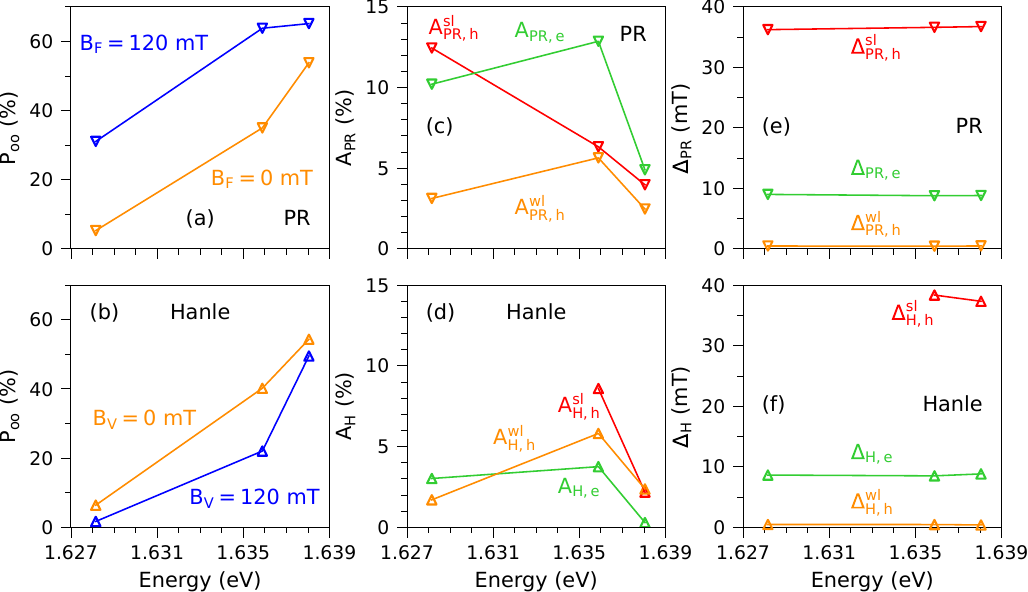}}
\caption{Spectral dependence of the parameters evaluated from the polarization recovery and Hanle curves, measured at $T=1.6$~K. The excitation laser helicity is modulated at 400~kHz, with $E_{\rm exc} = 1.722$~eV and $P = 18$~W/cm$^2$. (a,b) Optical orientation degree measured at zero field (orange) and 120~mT (blue). (c-f) Amplitudes and half-widths of the respective components for strongly localized holes (red),  electrons (green), and weakly localized holes (orange). In all panels, the symbols give the experimental data and the lines are guides for the eye.} 
\label{fig:Hanle_PRC_Spectral_Parameters}
\end{figure*}

\section{Spectral dependence of optical orientation: Hanle and polarization recovery}
\label{sec:Hanle}

There are two effects in magnetic field, which allow one to record comprehensive information on the carrier spins from optical orientation experiments. In transverse magnetic field (Voigt geometry), the optical orientation degree decreases due to the Hanle effect~\cite{meier1984optical}. In this case, the oriented carrier spins undergo Larmor precession about the magnetic field direction, which reduces the projection of their spin polarization $S_z$ onto the light propagation direction and decreases the optical orientation degree. In its classical form, the Hanle effect is expressed by the following equation:
\begin{equation}
S_{z}(B_{\rm V})=\frac{S_{z}(0)}{1+(B_{\rm V}/\Delta_{\rm H})^2}.
\label{eq:Hanle_formula}
\end{equation}
Here, $S_{z}(0)$ represents the spin projection along the initial spin polarization direction ($z$-axis) at $B_{\rm V}=0$. In fitting the experimental results, $S_{z}(0)$ corresponds to the amplitude of the Hanle curve, $A_{\rm H}$, which is the optical orientation degree at zero field. $B_{\rm V}$ is the magnetic field strength applied in the Voigt geometry. The characteristic field $\Delta_{\rm H}$ corresponds to the half-width at half-maximum (HWHM) of the Hanle curve. 

In contrast, when the magnetic field is applied along the direction of optical excitation (Faraday geometry), the relaxation of carrier spins induced by their interaction with the random hyperfine fields of the nuclear spin fluctuations is suppressed. This occurs because the carrier spin orientation is stabilized along the field direction. As a result, the optical orientation of carriers increases with growing field. This effect is known as polarization recovery (PR)~\citep{meier1984optical,smirnov2020spin}. The Hanle and PR effects were demonstrated experimentally and analyzed theoretically for lead halide perovskite semiconductors, e.g., in FA$_{0.9}$Cs$_{0.1}$PbI$_{2.8}$Br$_{0.2}$ crystals~\cite{kudlacik2024optical,Kotur2026_FAPI}.


The Hanle and polarization recovery curves measured at two detection energies of $E_{\rm det}=1.638$~eV and 1.628~eV are shown in Fig.~\ref{fig:Hanle_PRC_Spectral}. The shape of these curves is complex, indicating contributions from several processes. Qualitatively, the experimental features for MAPbI$_3$ are similar to those observed for FA$_{0.9}$Cs$_{0.1}$PbI$_{2.8}$Br$_{0.2}$~\cite{kudlacik2024optical,Kotur2026_FAPI}. Therefore, we apply the same assignment of the observed features to either electrons or holes. This assignment is further confirmed by the dynamic nuclear polarization experiments reported in Sec.~\ref{sec:DNP}.

The Hanle curves are fitted with the sum of three Lorentzian functions (Eq.~\eqref{eq:Hanle_formula}), resulting in the distinct individual contributions shown by the dashed lines in Fig.~\ref{fig:Hanle_PRC_Spectral}. The PR curves are fitted similarly, using as well the sum of three Lorentzian functions~\citep{kudlacik2024optical}
\begin{equation}
P_{\rm oo}(B_{\rm F})=A_{\rm max}-A_{\rm PR}\frac{\Delta_{\rm PR}^2}{B_{\rm F}^2+\Delta_{\rm PR}^2}.
\label{eq:PR_formula}
\end{equation}
Here, $A_{\rm max}$ is the optical orientation degree in the limit of an infinite Faraday field, $A_{\rm PR}$ is the amplitude and $\Delta_{\rm PR}$ is the HWHM of the PR curve. The parameters extracted from the fits of the Hanle and PR curves are given in Table~\ref{tab:comparison_spectral} and shown in Fig.~\ref{fig:Hanle_PRC_Spectral_Parameters}. 

Three features, which differ in their widths, are identified as resulting from: (i) strongly localized holes with $\Delta^{\rm sl}_{\rm H(PR),h}=37$~mT, (ii) localized electrons with $\Delta_{\rm H(PR),e}=9$~mT, and (iii) weakly localized holes with $\Delta^{\rm wl}_{\rm H(PR),h}=0.5$~mT. It is important to note that by strong or weak localization, we first of all consider the difference between smaller or larger localization volume, respectively. The carrier localization volume determines the number of nuclear spins interacting with the carrier within this volume~\cite{Meliakov2024,Meliakov2026} and, therefore, determines the widths of the Hanle and PR curves.

One can see in Fig.~\ref{fig:Hanle_PRC_Spectral}(a) that in the case of the Hanle effect detected at $E_{\rm det}=1.638$~eV, i.e., close to the exciton resonance, the optical orientation degree in the field range up to 125~mT does not drop to zero, but remains at about 50\%. This background polarization is assigned to the optical orientation of short-lived excitons, which we studied in detail in MAPbI$_3$ crystals in Ref.~\cite{kopteva2025optical}. Indeed, this background is absent at $E_{\rm det}=1.628$~eV, where the exciton contribution to the emission vanishes (Fig.~\ref{fig:Hanle_PRC_Spectral}(b)). The diminishing exciton contribution with decreasing $E_{\rm det}$ can be seen in the spectral dependence of $P_{\rm oo}(B_{\rm V}=120~{\rm mT})$, as shown by the blue symbols and the line in Fig.~\ref{fig:Hanle_PRC_Spectral_Parameters}(b). Hence, at detection energies close to the exciton resonance, the Hanle curve exhibits contributions not only from localized carriers but also from excitons. The exciton contribution gradually decreases and eventually vanishes with lowering detection energy.

\begin{table*}
\caption{Parameters of the Hanle and PR curves measured for MAPbI$_3$ crystals at different detection energies for $T=1.6$~K. For comparison, the parameters for a  FA$_{0.9}$Cs$_{0.1}$PbI$_{2.8}$Br$_{0.2}$ crystal from Ref.~\onlinecite{Kotur2026_FAPI} are given.
}
\label{tab:comparison_spectral}
\begin{center}
\begin{tabular*}{0.845\textwidth}{cccccccccc}
\hline
Sample & $E_{\rm X}$ (eV) & Curve & $E_{\rm det}$ (eV) & $A_{\rm h}^{\rm sl}$ (\%) & $A_{\rm e}$ (\%) & $A_{\rm h}^{\rm wl}$ (\%) & $\Delta_{\rm h}^{\rm sl}$ (mT) & $\Delta_{\rm e}$ (mT) & $\Delta_{\rm h}^{\rm wl}$ (mT) \\
\hline
\multirow{4}{*}{MAPbI$_3$} & \multirow{4}{*}{1.636} & \multirow{2}{*}{PR} & 1.638 & 4.0 & 4.9 & 2.5 & 36.7 & 8.8 & 0.5 \\
 &  &  & 1.628 & 12.5 & 10.2 & 3.1 & 36.2 & 9.0 & 0.5 \\ 
 &  & \multirow{2}{*}{Hanle} & 1.638 & 2.2 & 0.3 & 2.4 & 37.3 & 8.8 & 0.4 \\
 &  &  & 1.628 & $\approx 0$  & 3.0 & 1.7 & - & 8.6 & 0.5 \\ \hline
\multirow{4}{*}{FA$_{0.9}$Cs$_{0.1}$PbI$_{2.8}$Br$_{0.2}$} & \multirow{4}{*}{1.506} & \multirow{2}{*}{PR} & 1.504 & 2.6 & 4.6 & 5.8 & 19.1 & 5.0 & 0.4 \\
 &  &  & 1.498 & 1.4 & 1.7 & 2.0 & 16.6 & 4.5 & 0.3 \\ 
 &  & \multirow{2}{*}{Hanle} & 1.504 & 2.0 & 2.7 & 6.8 & 23.6 & 4.3 & 0.4 \\
 &  &  & 1.498 & 0.4 & 0.7 & 1.7 & 21.6 & 3.6 & 0.3 \\ \hline 
\end{tabular*}
\end{center}
\end{table*}

The spectral dependences of the HWHM of the PR and Hanle curves are shown in Figs.~\ref{fig:Hanle_PRC_Spectral_Parameters}(e) and \ref{fig:Hanle_PRC_Spectral_Parameters}(f). One can see that they do not show any dispersion, having constant values of about 37, 9, and 0.5~mT. It is important to note that the widths are the same for the PR and the Hanle curves for each carrier type. From this fact and based on the model analysis in Ref.~\onlinecite{kudlacik2024optical} (see Fig.~2(d) there), we conclude that in the MAPbI$_3$ crystals the carrier correlation time with the nuclear spin fluctuations, $\tau_{\rm c}$, exceeds the carrier spin lifetime, $T_{\rm s}$. The correlation time is contributed by either carrier hopping between two nuclear fluctuations (i.e., between two localized states) or orientation change of the nuclear spin fluctuation in the same localized state. In semiconductors, the second process takes typically much longer than the first one.  In the regime of $T_{\rm s} < \tau_{\rm c}$, the ratio of the PR and the Hanle amplitude should be $A_{\rm PR}/A_{\rm H}=3/1$. However, the experimental data shown in Fig.~\ref{fig:Hanle_PRC_Spectral_Parameters}(c) and \ref{fig:Hanle_PRC_Spectral_Parameters}(d) reveal a deviation from the expected ratio. The reason is that the experimentally measured Hanle and PR curves contain three or four contributions, including excitons, making it difficult to extract the true amplitudes. A complete analysis would require also knowledge of the relative intensities of the emission processes, that is currently not available. Therefore, in our analysis, we rely on the widths, which can be reliably determined from the experiment.


The width values are comparable with the corresponding HWHM measured in Ref.~\cite{Kotur2026_FAPI} for FA$_{0.9}$Cs$_{0.1}$PbI$_{2.8}$Br$_{0.2}$ crystals: 20, 4, and 0.4~mT.  Note, e.g., that for the strongly localized holes, $\Delta^{\rm sl}_{\rm H(PR),h}=37$~mT in MAPbI$_3$ whereas in FA$_{0.9}$Cs$_{0.1}$PbI$_{2.8}$Br$_{0.2}$ it has about half the value, $\Delta^{\rm sl}_{\rm H(PR),h}=20$~mT. However, to evaluate the number of Pb nuclear spins interacting with the hole ($N_L$), the hole $g$-factor must also be taken into account~\cite{Meliakov2024}. They are $g_{\rm h}=-0.54$ in MAPbI$_3$ and $-1.15$ in FA$_{0.9}$Cs$_{0.1}$PbI$_{2.8}$Br$_{0.2}$, i.e., they differ by about a factor of two. Thus, the differences compensate each other, yielding approximately the same $N_L$ or hole localization volume in both materials.

\section{Temperature dependence of optical orientation}
\label{sec:temperature}

The temperature dependences of the Hanle and PR signals provide further insight into the spin dynamics of electrons and holes interacting with the nuclear spins. We measure them in the temperature range of $1.6-15$~K, where the optical orientation degree is sufficiently large to be assessed.

\begin{figure*}[hbt]
\center{\includegraphics{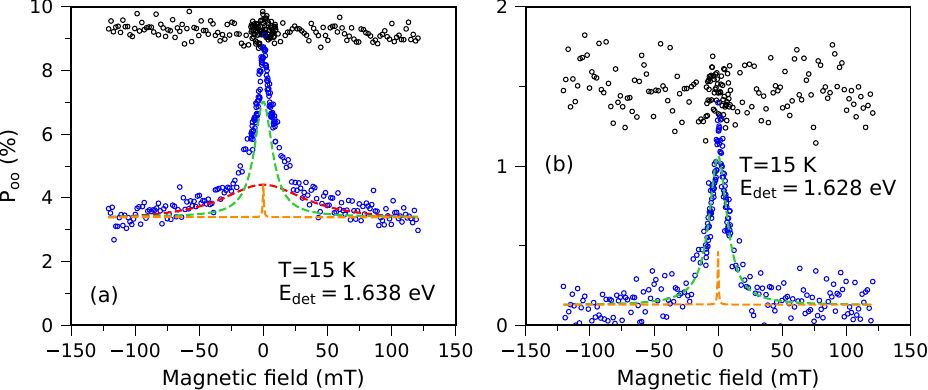}}
\caption{Polarization recovery and Hanle curves measured at $T = 15$~K with $\sigma^+/\sigma^-$ excitation modulated at 400~kHz for two detection energies, (a) $E_{\rm det} = 1.638$~eV and (b) $E_{\rm det} = 1.628$~eV. $P = 18$~W/cm$^2$ and $E_{\rm exc} = 1.722$~eV. The dashed lines show the individual contributions of strongly localized holes (red), electrons (green), and weakly localized holes (orange).}  
\label{fig:Hanle_PRC_Temperature}
\end{figure*}

In Figure~\ref{fig:Hanle_PRC_Temperature}, the Hanle and PR curves measured at $T=15$~K are shown for detection energies of $E_{\rm det}=1.638$~eV and $1.628$~eV. At the exciton energy of 1.638~eV, the optical orientation degree at zero magnetic field is reduced from about 54\% at $1.6$~K (Fig.~\ref{fig:Hanle_PRC_Spectral}(a)) to about 9\% at $15$~K, see Fig.~\ref{fig:Hanle_PRC_Temperature}(a). At the energy of $1.628$~eV the decrease is from about 6\% (Fig.~\ref{fig:Hanle_PRC_Spectral}(b)) to 1.5\%, see Fig.~\ref{fig:Hanle_PRC_Temperature}(b). These values at 15~K are insensitive to the Faraday magnetic field strength, meaning that the PR curves have zero amplitude then. This evidences carrier delocalization at elevated temperatures. The Hanle curves are still pronounced, albeit with considerably reduced amplitudes.

\begin{table*}
\caption{Parameters of the Hanle and PR curves measured for a MAPbI$_3$ crystal for different temperatures at a detection energy of 1.628~eV.}
\label{tab:comparison_temperature}
\begin{center}
\begin{tabular*}{0.56\textwidth}{ccccccccc}
\hline
 Curve & $T$ (K) & $A_{\rm h}^{\rm sl}$ (\%) & $A_{\rm e}$ (\%) & $A_{\rm h}^{\rm wl}$ (\%) & $\Delta_{\rm h}^{\rm sl}$ (mT) & $\Delta_{\rm e}$ (mT) & $\Delta_{\rm h}^{\rm wl}$ (mT) \\
\hline
\multirow{4}{*}{PR} & 1.6 & 12.5 & 10.2 & 3.1 & 36.2 & 9.0 & 0.5 \\
  & 6 & 9.7 & 2.8 & 1.7 & 41.4 & 10.5 & 0.4 \\ 
  & 10 & 2.8 & 0.7 & 0 & 43.8 & 10.9 & - \\ 
  & 15 & 0 & 0 & 0 & - & - & - \\ \hline
 \multirow{4}{*}{Hanle} & 1.6 & $\approx 0$ & 3.0 & 1.7 & - & 8.6 & 0.5 \\ 
  & 6 & $\approx 0$ & 1.9 & 1.6 & - & 9.0 & 0.4 \\ 
  & 10 & $\approx 0$ & 1.2 & 1.0 & - & 8.5 & 0.4 \\
  & 15 & $\approx 0$ & 0.9 & 0.3 & - & 8.5 & 0.4 \\ \hline
\end{tabular*}
\end{center}
\end{table*}

\begin{figure*}[hbt]
\center{\includegraphics{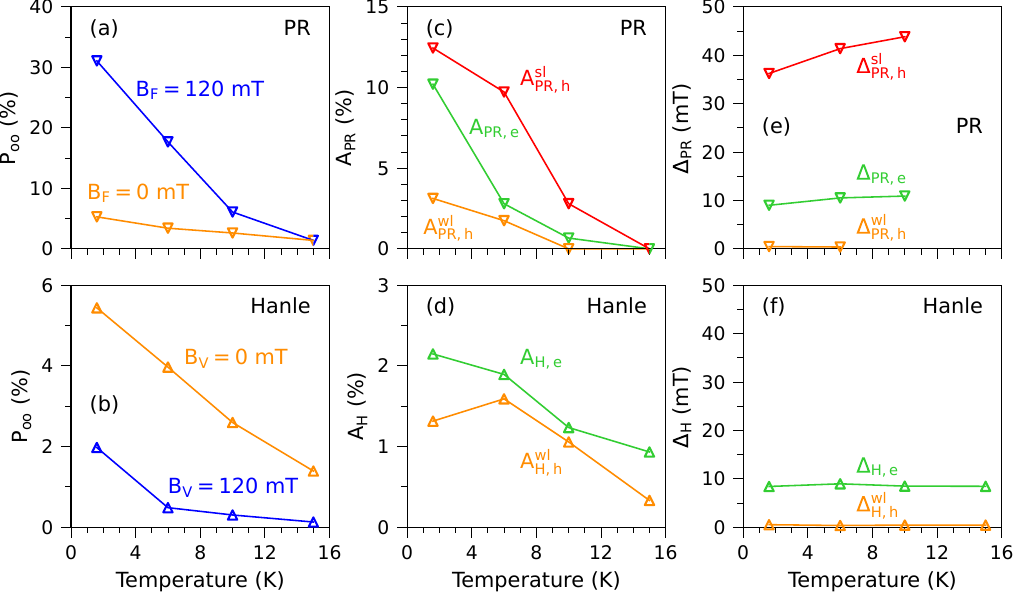}}
\caption{Temperature dependence of the parameters evaluated from the polarization recovery and Hanle curves measured at $E_{\rm det}=1.628$~eV. The circular laser helicity is modulated at 400~kHz. $E_{\rm exc}=1.722$~eV and $P=18$~W/cm$^2$. (a,b) Total optical orientation degree measured at zero field (orange) and 120~mT (blue). (c-f) Amplitudes and HWHMs of the respective components of strongly localized holes (red), electrons (green), and weakly localized holes (orange). In all panels, the symbols give the experimental data and the lines are guides to the eye.}
\label{fig:Hanle_PRC_Temperature_Parameters}
\end{figure*}

The evaluated parameters describing the charge carrier contributions at four different temperatures for $E_{\rm det}=1.628$~eV are given in Table~\ref{tab:comparison_temperature}. Their temperature dependences are shown in Fig.~\ref{fig:Hanle_PRC_Temperature_Parameters}. For the $P_{\rm oo}$ values we present data at 0 and 120~mT, whose difference represents the amplitude of the PR (Fig.~\ref{fig:Hanle_PRC_Temperature_Parameters}(a)) and Hanle (Fig.~\ref{fig:Hanle_PRC_Temperature_Parameters}(b)) signals. The temperature dependences of the decomposed carrier amplitudes are shown in Figs.~\ref{fig:Hanle_PRC_Temperature_Parameters}(c,d).

For temperatures between 1.6 and 10~K, the PR curves exhibit three contributions with characteristic HWHMs of about 40, 9, and 0.5~mT (Fig.~\ref{fig:Hanle_PRC_Temperature_Parameters}(e)). The Hanle curves contain only two contributions with HWHMs of about 9 and 0.5~mT (Fig.~\ref{fig:Hanle_PRC_Temperature_Parameters}(f)). The widest contribution, attributed to strongly localized holes, is not visible at this energy, presumably because its amplitude is below the detection sensitivity.

At temperatures above 10~K, the PR curve can no longer be observed. A similar behavior was found for FA$_{0.9}$Cs$_{0.1}$PbI$_{2.8}$Br$_{0.2}$ crystals, where the PR amplitude dropped to zero at temperatures exceeding 8~K~\cite{Kotur2026_FAPI}. This behavior was attributed to thermal delocalization of charge carriers at elevated temperatures, which diminishes the role of nuclear spin fluctuations in the carrier spin relaxation.

For temperatures between 10 and 15~K, the full amplitude of the Hanle curve decreases, but remains nonzero, see the orange symbols and the line in Fig.~\ref{fig:Hanle_PRC_Temperature_Parameters}(b). In this temperature range, the PR curve disappears, indicating that the relaxation channel of carrier spins mediated by nuclear spin fluctuations is suppressed as result of carrier delocalization. This indicates that for temperatures above 10~K, the carrier spin relaxation is determined by their intrinsic spin lifetime $T_{\rm s}$, directly reflecting the Hanle width. In this case, the Hanle effect can be described by Eq.~\eqref{eq:Hanle_formula}, where the characteristic field $\Delta_{\rm H, e(h)}$ is inversely proportional to the spin lifetime $T_{\rm s,e(h)}$ of electrons (holes)
\begin{equation}
\Delta_{\rm H,e(h)}=\frac{\hbar}{\mu_{\rm B} g_{\rm e(h)} T_{\rm s,e(h)}}.
\label{eq:Ts_high_T}
\end{equation}
Here $\hbar$ is the reduced Planck constant, $\mu_{\rm B}$ the Bohr magneton, and $g_{\rm e(h)}$ the electron (hole) $g$-factor. The spin lifetime is determined by
\begin{equation}
\frac{1}{T_{\rm s,e(h)}}=\frac{1}{\tau_{\rm r,e(h)}}+\frac{1}{\tau_{\rm s,e(h)}},
\end{equation}
where $\tau_{\rm r,e(h)}$ is the recombination time and $\tau_{\rm s,e(h)}$ the spin relaxation time of electrons (holes). At $E_{\rm det} = 1.638$~eV for $T=15$~K the Hanle HWHM for electrons and holes are about 10~mT and 45~mT, respectively, see Fig.~\ref{fig:Hanle_PRC_Temperature}(a). With the corresponding $g$-factors of $g_{\rm e} = +2.83$ and $g_{\rm h} = -0.54$~\citep{kopteva2025optical} in combination with Eq.~\eqref{eq:Ts_high_T}, this provides an evaluation of the spin lifetimes of $T_{\rm s,e} =400$~ps and $T_{\rm s,h} = 470$~ps.

\section{Dynamic polarization of nuclear spins}
\label{sec:DNP}


Dynamic nuclear polarization (DNP) is induced by the transfer of spin angular momentum from optically oriented carriers to the lattice nuclei via the hyperfine interaction~\citep{meier1984optical}. For dynamic polarization and its accumulation, the nuclear spins have to be exposed to an external magnetic field, that is non-orthogonal to the carrier mean spin. 
In addition, the circularly polarized excitation driving the non-equilibrium carrier spin polarization must have a fixed helicity, either $\sigma^+$ or $\sigma^-$. Through DNP, the effective Overhauser magnetic field $B_{\rm N}$ of the spin-polarized nuclei develops. Depending on the helicity of the excitation light, the Overhauser field can be parallel or antiparallel to the external magnetic field. As a result, the optically oriented carrier spins are subject to the total field $\textbf{B}_{\rm tot}=\textbf{B} \pm \textbf{B}_{\rm N}$.

\begin{figure*}[hbt]
\center{\includegraphics{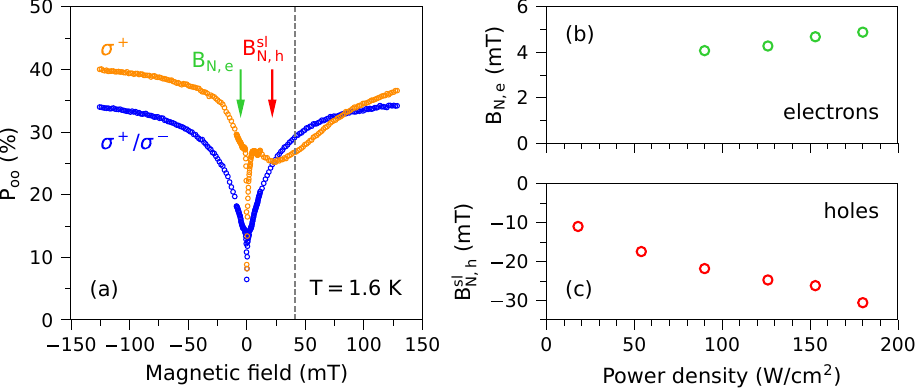}}
\caption{(a) Polarization recovery curves measured in Faraday geometry for modulated $\sigma^+/\sigma^-$ (blue) and fixed $\sigma^+$ (orange) polarized excitation at $T = 1.6$~K. $E_{\rm exc}=1.722$~eV, $P=90$~W/cm$^2$, and $E_{\rm det}=1.628$~eV. The arrows indicate the magnitudes of the Overhauser fields for electrons $B_{\rm N,e}$ and holes $B_{\rm N,h}^{\rm sl}$. The dashed gray line indicates the external magnetic field applied during the measurement of the nuclear spin build-up time $T_{\rm 1,N}^{\rm light}$ under optical pumping shown in Fig.~\ref{fig:Relaxation}. The dependence of the Overhauser field on excitation density is shown for electrons in panel (b) and for holes in panel (c).}
\label{fig:PRC_DNP}
\end{figure*}

We use the Faraday geometry, where the external magnetic field is aligned with the light propagation direction. Excitation is provided by circularly polarized light at an energy of $E_{\rm exc} = 1.722$~eV with modulated ($\sigma^+/\sigma^-$) or fixed ($\sigma^+$) helicity. The resulting polarization recovery curves for an excitation density of 90 W/cm$^2$ are shown in Fig.~\ref{fig:PRC_DNP}(a). The symmetric PR curve is obtained under modulated $\sigma^+/\sigma^-$ excitation, i.e., for conditions where DNP is not expected. For constant $\sigma^+$ excitation, a pronounced asymmetry of the PR curve appears. It is characterized by shoulders centered at $B_{\rm N,h}^{\rm sl} \approx -22$~mT and $B_{\rm N,e} \approx 4$~mT. The shoulder minimum occurs when the particular Overhauser field is compensated by the external magnetic field, i.e., $B_{\rm N}=-B$.  This explains the negative sign of $B_{\rm N}$ for the shift to positive external fields. The opposite signs of $B_{\rm N,e}$ and $B_{\rm N,h}^{\rm sl}$, which are evident in Fig.~\ref{fig:PRC_DNP}, are related to the opposite signs of the electron and hole $g$-factors in MAPbI$_3$ crystals: $g_{\rm e}=+2.83$ and $g_{\rm h}=-0.54$~\cite{kopteva2025optical}. Note that in lead halide perovskites, the hyperfine interaction of electrons with the nuclei is significantly weaker than that of the holes, resulting in a comparatively smaller Overhauser field. On this basis, the Overhauser field of $B_{\rm N,h}^{\rm sl}=-22$~mT has to be attributed to created by lattice nuclei polarized through the hyperfine interaction with holes, whereas the field of $B_{\rm N,e}=4$~mT is attributed to the interaction of nuclei with electrons.


The excitation density dependence of the Overhauser fields acting on the electrons and the holes are shown in Figs.~\ref{fig:PRC_DNP}(b,c), covering the range of $P$ from 18 to 180~W/cm$^2$. 
It is evident that stronger excitation induces larger Overhauser fields, as more spin-polarized carriers are generated, resulting in a larger nuclear spin polarization. For the largest excitation density, the Overhauser field acting on the electrons reaches $5$~mT, while that acting on the holes reaches $-30$~mT.

For comparison, the Overhauser fields reported for FA$_{0.9}$Cs$_{0.1}$PbI$_{2.8}$Br$_{0.2}$ crystals are  $B_{\rm N,e}=+11$~mT  and $B_{\rm N,h}^{\rm sl}=-59$~mT for a laser power density of 88~W/cm$^2$~\cite{Kotur2026_FAPI}. For the same excitation conditions, the Overhauser fields in the MAPbI$_3$ crystal are three times weaker. This may explain the fact that the DNP shift of the narrow curve corresponding to weakly localized holes is not observed in MAPbI$_3$, whereas it reaches $B_{\rm N,h}^{\rm wl} = -1.4$~mT in FA$_{0.9}$Cs$_{0.1}$PbI$_{2.8}$Br$_{0.2}$~\cite{Kotur2026_FAPI}. 


\begin{figure}[hbt]
\center{\includegraphics{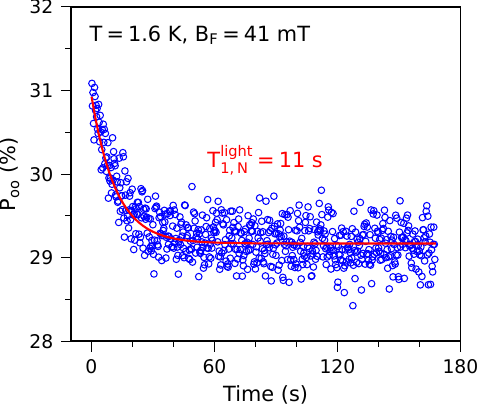}}
\caption{Dynamic build-up of nuclear spin polarization probed through the change of $P_{\rm oo}$ following a dark interval of $t_{\rm dark}=200$~s at $B_{\rm F}=41$~mT. The red line represents a monoexponential fit to the data, giving a nuclear spin build-up time under illumination of $T_{\rm 1,N}^{\rm light}=11$~s. $T=1.6$~K. $E_{\rm exc} = 1.722$~eV, $P = 90$~W/cm$^2$, and $E_{\rm det} = 1.628$~eV.}
\label{fig:Relaxation}
\end{figure}

The build-up time of the nuclear spin polarization is measured in a magnetic field of $B_{\rm F}=41$~mT applied in the Faraday geometry. To allow the nuclear spins to relax, the excitation laser was blocked for a dark interval of $t_{\rm dark}=200$~s. After unblocking the laser, the time evolution of the optical orientation degree $P_{\rm oo}(t)$ is recorded, which reflects the build-up of the nuclear spin polarization induced by the DNP. Figure~\ref{fig:Relaxation} shows the dynamics of $P_{\rm oo}(t)$, which demonstrates a gradual decrease of the optical orientation degree followed by saturation. This process reflects the evolution from the dependence shown by the blue circles (no DNP) in Fig.~\ref{fig:PRC_DNP}(a) to that shown by the orange circles (with DNP), at the magnetic field indicated by the dashed line.
The time-dependent behavior of the Overhauser field follows the dependence~\citep{mocek2017high}
\begin{equation}
B_{\rm N}(t)=B_{\rm N}^{\rm st}[1-\exp({-t/T_{\rm 1,N}})],
\end{equation}
where $B_{\rm N}^{\rm st}$ is the steady-state Overhauser field. Fitting the measured $P_{\rm oo}(t)$ dynamics in Fig.~\ref{fig:Relaxation} to an exponential decay $\sim \exp(-t/T_{\rm 1,N})$ gives the build-up time of the nuclear spin polarization under illumination of $T_{\rm 1,N}^{\rm light}=11$~s. This value is close to the $T_{\rm 1,N}^{\rm light}=9$~s measured for FA$_{0.9}$Cs$_{0.1}$PbI$_{2.8}$Br$_{0.2}$ crystals~\citep{Kotur2026_FAPI}.

\section{Conclusions}

Our experimental results show that in MAPbI$_3$ single crystals the spin dynamics of localized electrons and holes at cryogenic temperatures are controlled by the hyperfine interaction with the nuclear spin system. The complex shapes of the Hanle and the polarization recovery curves measured in the optical orientation experiments under continuous wave excitation reveal the contributions of localized electrons, as well as strongly and weakly localized holes. 
Also, the localization volumes for carriers are comparable in both materials, which shows that the alloy fluctuations inherent to FA$_{0.9}$Cs$_{0.1}$PbI$_{2.8}$Br$_{0.2}$ crystals do not determine the carrier localization conditions. The same values of the widths of the PR and the Hanle curves for the respective carriers indicate that in MAPbI$_3$ crystals the carrier correlation time with the nuclear spin fluctuations exceeds the carrier spin lifetime, $T_{\rm s}<\tau_{\rm c}$. Dynamic nuclear polarization induced by spin transfer from oriented electrons and holes is demonstrated for the MAPbI$_3$ crystals. The measured nuclear polarization dynamics shows a characteristic build-up time of $T_{\rm 1,N}^{\rm light}=11$~s.





\section*{Acknowledgments} 
The authors are thankful to N. E. Kopteva, K. V. Kavokin, and P. S. Bazhin  for fruitful discussions. M.K. acknowledges support by the BMBF project QR.N (Contract No.16KIS2201). D.R.Y. acknowledges support by the Deutsche Forschungsgemeinschaft via the SPP2196 Priority Program (Project YA 65/28-1, No. 527080192).  The work at ETH Z\"urich (B.T. and M.V.K.) was financially supported by the Swiss National Science Foundation (grant agreement 186406, funded in conjunction with SPP219 through DFG-SNSF bilateral program) and by ETH Z\"urich through ETH+ Project SynMatLab.

\textbf{ORCID }\\
Mladen Kotur:        0000-0002-2569-5051 \\  
Nataliia E. Kopteva: 0000-0003-0865-0393 \\  
Dmitri R. Yakovlev:  0000-0001-7349-2745 \\  
Bekir Turedi:         0000-0003-2208-0737 \\ 
Maksym V. Kovalenko:  0000-0002-6396-8938 \\ 
Manfred Bayer        0000-0002-0893-5949 \\  


\end{document}